\documentclass[a4paper, aps, amsmath, 10pt, nofootinbib, superscriptaddress]{revtex4-2}

\usepackage{graphicx}
\usepackage{anyfontsize}
\usepackage{xcolor}
\usepackage{siunitx}
\usepackage{physics}
\usepackage{amsfonts}
\usepackage{mathtools}
\usepackage{svg}
\usepackage{braket}
\usepackage[skip=2pt,font=normalsize]{subcaption}
\usepackage{commath}
\usepackage{environ}
\usepackage{placeins}

\usepackage[english]{babel}
\usepackage{amsthm, amssymb, calrsfs, verbatim, bbm, bbold, color, graphics, slashed}

\usepackage{tikz}
\usepackage{tikz-3dplot}
\usepackage{tikz-cd}

\usetikzlibrary{arrows.meta}
\usetikzlibrary{shapes.misc}
\usetikzlibrary{shapes.geometric}
\usetikzlibrary{decorations.markings}
\usetikzlibrary{patterns}

\tikzset{
    crossed dot/.style={
        /tikzfeynman/crossed dot
    },
    photon/.style={
        /tikzfeynman/photon
    },
    scalar/.style={
    },
    cross/.style={
        cross out,
        draw=black,
        fill=none,
        minimum size=2*(#1-\pgflinewidth),
        inner sep=0pt,
        outer sep=0pt
    },
    cross/.default={2pt}
}

\usepackage[compat=1.1.0]{tikz-feynman}
\usepackage{hyperref}

\newtheorem*{eqtheorem}{The Equivalence Theorem}

\newcommand\at[2]{\left.#1\right|_{#2}}
\newcounter{feyndiag}

\NewEnviron{feyndiag}[1][]{
    \refstepcounter{feyndiag}
    \begin{equation}
        \begin{tikzpicture}[#1]
            \BODY
        \end{tikzpicture}
        \tag{D\thefeyndiag}
    \end{equation}
}

\NewDocumentCommand{\snail}{O{5cm} m O{#2} O{0.5cm}}{%
\parbox{#1}{
\begin{feyndiag}[baseline=(i)]
    \label{diag:snail-#2-#3};
    \coordinate (i) at (-1cm,0);
    \coordinate (o) at (1cm,0);
    \coordinate (v) at (0,0);
    \draw[dotted] ([xshift=-0.5cm]i) node[left] {\(J\)} -- (i) (o) -- ([xshift=0.5cm]o) node[right] {\(J^\dag\)};
    \draw[#2] (i) -- (v) -- (o);
    \draw[#3] (v) +(0,#4) circle[radius=#4];
\end{feyndiag}}
}
\NewDocumentCommand{\sunset}{O{5cm} m m O{#2}}{%
\parbox{#1}{
\begin{feyndiag}[baseline=(i)]
    \label{diag:sunset-#2-#3-#4};
    \coordinate (i) at (-1cm,0);
    \coordinate (o) at (1cm,0);
    \coordinate (v1) at (-0.5cm,0);
    \coordinate (v2) at (0.5cm,0);
    \draw[dotted] ([xshift=-0.5cm]i) node[left] {\(J\)} -- (i) (o) -- ([xshift=0.5cm]o) node[right] {\(J^\dag\)};
    \draw[#2] (i) -- (v1) (v2) -- (o);
    \draw[#3] (v1) arc[radius=0.5cm,start angle=180,end angle=0];
    \draw[#4] (v2) arc[radius=0.5cm,start angle=0,end angle=-180];
\end{feyndiag}}
}

\begin{document}

\title{The Equivalence Theorem at work: manifestly gauge-invariant Abelian Higgs model physics}

\author{Bram Boeykens}
\email{bramwannesboeykens@gmail.com}
\affiliation{KU Leuven Campus Kortrijk -- Kulak, Department of Physics, Etienne Sabbelaan 53 bus 7657, 8500 Kortrijk, Belgium}

\author{David Dudal}
\email{david.dudal@kuleuven.be}
\affiliation{KU Leuven Campus Kortrijk -- Kulak, Department of Physics, Etienne Sabbelaan 53 bus 7657, 8500 Kortrijk, Belgium}

\author{Thomas Oosthuyse}
\email{thomas.oosthuyse@kuleuven.be}
\affiliation{KU Leuven Campus Kortrijk -- Kulak, Department of Physics, Etienne Sabbelaan 53 bus 7657, 8500 Kortrijk, Belgium}

\begin{abstract}
    We reconsider the Equivalence Theorem from an algebraic viewpoint, using an extended BRST symmetry.  This version of the Equivalence Theorem is then used to reexpress the Abelian Higgs model action, originally written in terms of undesirable gauge variant field excitations, in terms of gauge-invariant, physical variables, corresponding to the Fr\"ohlich--Morchio--Strocchi composite operators in the original field formulation. Although the ensuing action encompasses an infinite number of vertices and appears to be nonrenormalizable from the powercounting viewpoint, it nevertheless is renormalizable, thanks to the hidden equivalence with the original model. Hence, manifestly gauge-invariant computations are possible.  We present an explicit illustration in terms of the gauge-invariant scalar field, its Green's function and corresponding pole mass.
\end{abstract}

\maketitle

\section{Introduction}
In many areas of physics and mathematics, the initial variables may prove suboptimal for efficient calculations or to clearly describe the physical content of the model. A change of variables may sometimes provide an easier calculation or lead to clearer insights. In classical theories it is well known that such changes of variables do not influence physics, thinking for example about canonical transformations in Hamiltonian mechanics.

In quantum field theory, a change of variables is not that straightforward, given the fact that physical observables are usually determined from scattering processes and thus $\cal S$-matrix elements, which in quantum field theory, are to be computed from correlation functions. In perturbation theory, this happens via Feynman diagrams. From this perspective, non-linear field transformations might interfere with the very concept of a free field Lagrangian around which ones perturbs, or even complicate the local description of the theory. Also renormalizability might become clouded if the field variables are chosen unwisely. Nonetheless, it was already discussed in \cite{eqtheoremchisholm} that field transformations leaving the free field Lagrangian invariant should not alter the $\cal S$-matrix, this became commonly known as The Equivalence Theorem. Deeper takes at this theorem can be found in e.g.~\cite{Kamefuchi:1961sb,Bergere:1975tr,eqtheorem_blasi,Ferrari:2002kz,Cohen:2024fak}. In particular did \cite{eqtheorem_blasi} provide a BRST symmetry perspective. In the current paper, we will refine the analysis of \cite{eqtheorem_blasi}, see also \cite{Ferrari:2002kz}, to allow for a more clean identification of which field operators are the physical ones by deriving a Nielsen-like identity \cite{Nielsen:1975fs,Piguet:1984js} in relation to the non-linear part of the transformations.  As our interest is mostly in gauge theories, with ``physical'' we then mean operators that are both BRST invariant and renormalizable. Our goal is to reformulate the action of gauge theories with a Higgs mechanism in terms of explicitly gauge (BRST) invariant operators as new basic field operators,  without spoiling the good UV properties, that is, renormalizability must be ensured at the level of correlation functions of the new operators. In relation to the Equivalence Theorem, this means that the admissible, non-linear field transformation that we may consider can only be those mapping the original fields to renormalizable (composite) operators of these original fields. This is an essential new ingredient into the construction worked out in this paper, on top of the theorem itself.  

In the second section, we will first give an algebraic proof of the Equivalence Theorem using an extended BRST symmetry. In the third section we will apply our version of the Equivalence Theorem to the Abelian Higgs model, continuing the work of \cite{Dudal2021-md}.  We will consider the relevant  FMS (Fr\"ohlich--Morchio--Strocchi) operators \cite{fmspaper,fmspaper2,fmsreviewmaas} to capture the physical spectrum in a gauge-invariant fashion. The need for such operators was already realized earlier by 't Hooft in \cite{gerard1980recent}. Finally, we will perform a change of variables to obtain a renormalizable theory in terms of exactly these FMS operators. An illustration in terms of the gauge-invariant scalar is worked out in the fourth section, before ending with our conclusions. Various technical details are collected in the Appendices.

\section{The Equivalence Theorem itself from a BRST viewpoint}
Suppose we start from an initial action $S_0[\phi]$, where the physical, renormalizable (possibly composite) operators of interest are given by $T_i(x)=F(\phi,\partial_\mu \phi,\dots)$, in other words, built from the fields $\phi$ and its derivatives. We can obtain correlation functions of these $T_i$ by taking functional derivatives of the path integral, introducing the external sources $Q_i$ in the action from the start:
\begin{equation}
	S_\text{ext}[\phi,Q]=S_0[\phi]+\int \dd[4]{x} Q_i(x) T_i(x).
\end{equation} 
Then the following holds:
\begin{eqtheorem}
	Correlation functions, between the operators $T_i$ are independent on a local and invertible change of variables of the form $\phi[\hat\phi] = \hat{\phi}+ \hat{\phi}^2 g(\hat{\phi})$.
\end{eqtheorem}
Importantly, the operators $T_i$ transform also with the changes of variables. The transformed operator $T_i$ is simply given by the operator coupling to $Q_i$, but rewritten in terms of the new variables, that is,
\[T_i(x)=F(\phi,\partial_\mu \phi,\dots)\longrightarrow F(\phi[\hat\phi],\partial_\mu \phi[\hat\phi],\dots).\]

\subsection{BRST proof of the Equivalence Theorem}\label{eqtheorem}
The goal is to obtain a new theory, written in terms of the (new) field excitations $\hat{\phi}$, afterwards we are interested in how correlation functions between both formulations relate to one another. Crucial here is to realize that the Equivalence theorem does \emph{not} imply that correlation functions of $\hat{\phi}$ itself will be renormalizable or physically meaningful.

To start, we will perform the following field redefinition implicitly:
\begin{equation}\label{trans0}
	\phi \longrightarrow \hat{\phi} + \alpha \hat{\phi}^2 g(\alpha;\hat{\phi}).
\end{equation}
Here, $\alpha$ is a fictitious parameter that we introduced to get the correct mass dimensionality, whilst it will also allow to gain control over the non-linear part of the transformation \eqref{trans0}. Tacitly, we assume the $\alpha$-dependence to be analytical.

Next, we insert the following identity in the path integral, similarly to a Faddeev-Popov gauge fixing, where $N$ is an  irrelevant normalization factor,
\begin{equation}
	1=N\int [D\hat{\phi}]\det(1+ 2\alpha \hat{\phi} g +\alpha \hat{\phi}^2 \frac{\delta}{\delta \hat{\phi}}g) \delta(\hat{\phi}+\alpha \hat{\phi}^2g-\phi).
\end{equation}
Both contributions can be exponentiated into the action, using Grassmannian ghost fields $\bar{c}$ and $c$ for the Jacobian and a Lagrangian multiplier field $\chi$ for the $\delta$-constraint, whence
\begin{align*}
	Z[Q_i]&=N\int [D\phi][D\hat{\phi}][D\chi][Dc][D\bar{c}]\exp{-S_\text{tot}[\phi,\hat{\phi},\chi,c,\bar{c}]}
\end{align*}
with the action
\begin{equation}
\begin{aligned}
	S_\text{tot}[\phi,\hat{\phi},\chi,c,\bar{c}]	&=S_\text{ext}[\phi,Q_i]+\int \dd[4]{x} \Bigg(i\chi(\hat{\phi}+\alpha \hat{\phi}^2g-\phi)+\bar{c}\Big(1+2\alpha \hat{\phi} g +\alpha \hat{\phi}^2 \frac{\delta}{\delta \hat{\phi}}g\Big)c\Bigg).
\end{aligned}
\end{equation}
Functional integration over \(\chi\) and \(\phi\) then effectively performs the substitution \eqref{trans0}, but for the following analysis we keep the action as-is.
We would like for \(S_\text{tot}\) to obey a BRST like symmetry, as we then may characterize the physical subspace using BRST tools and use the implications of this symmetry on the structure of the quantum effective action and spectrum. We will also allow the parameter $\alpha$ to vary following \cite{Piguet:1984js}. The desired action of the infinitesimal generator on the fields is given by:
\begin{equation}
	\delta \phi=0\,,\quad \delta \hat{\phi}=c\,,\quad \delta c=0\,,\quad \delta\bar{c}=i \chi\,,\quad \delta\chi=0\,,\quad \delta \alpha =\beta\,,\quad \delta\beta=0.
\end{equation}
Its nilpotency, $\delta^2=0$, can easily be verified, moreover when we add the term $\bar{c}\beta \hat{\phi}^2g$ to the action, the complete action becomes invariant under this BRST-like symmetry.
\begin{equation}
    \begin{aligned}
    	S_\text{tot}[\phi,\hat{\phi},\chi,c,\bar{c},\beta] &= S_\text{ext}[\phi,Q] + \int \dd[4]{x} \qty[i\chi \qty(\hat{\phi}+\alpha \hat{\phi}^2g-\phi)+\bar{c}(1+\alpha \frac{\delta}{\delta \hat{\phi}} \hat{\phi}^2g)c+\bar{c}\beta \hat{\phi}^2g ]\\
        &=S_\text{ext}[\phi,Q]+ \delta\int \dd[4]{x} \bar{c} \qty(\hat{\phi}+\alpha \hat{\phi}^2g-\phi).
    \end{aligned}
\end{equation}
Although it seems that we have changed our theory by including the $\beta$ parameter, in the limit of $\beta\to 0$ we recover the original theory.
As a consequence the new action is simply an extension of the original action, in which we can more easily show that $\alpha$ does not influence correlation functions between $T_i$.
This can already be appreciated from the fact that $\alpha$ only appears in an $\delta$-exact term in the above action, akin to a gauge parameter when doing BRST quantization.

This (linear) BRST symmetry implies a Slavnov-Taylor identity on the quantum effective action,
\begin{equation}
	\int \dd[4]{x} \qty[c\frac{\delta}{\delta \hat{\phi}}+i \chi \frac{\delta}{\delta \bar{c}}+\beta \frac{\delta}{\delta \alpha}]\cdot \Gamma[Q_i]=0.
\end{equation}
A Legendre transformation of this expression yields a functional constraint on the generator of connected correlation functions,
\begin{equation}\label{eq_theorem_gen_func}
	\int \dd[4]{x} \qty[J_{\hat{\phi}}\frac{\delta}{\delta J_c}+iJ_{\bar{c}} \frac{\delta}{\delta J_\chi}+\beta \frac{\delta}{\delta \alpha}]\cdot Z^C[Q_i]=0.
\end{equation}
Finally, we can obtain correlation functions of the field operators $T_i$ by taking functional derivatives with respect to $Q$ as follows:
\begin{equation}
	\frac{\delta^n}{\delta Q_{i_1}(x_1)\dots \delta Q_{i_n}(x_n)} \at{Z^C[Q_i]}{J=0},
\end{equation}
or, applying such derivative to \eqref{eq_theorem_gen_func} and sending $J_{\hat{\phi}}$ and $J_{\bar{c}}$ to zero, gives us the desired Nielsen identity \cite{Piguet:1984js}
\begin{equation}\label{eqtheorem_exact}
	\beta \frac{\delta}{\delta \alpha}\frac{\delta^n}{\delta Q_{i_1}(x_1)\dots \delta Q_{i_n}(x_n)} \at{Z^C[Q_i]}{J=0}=0.
\end{equation}
Therefore, we obtain that functional derivatives with respect to the external source $Q_i(x)$ do not depend on the value of $\alpha$. Said otherwise, correlation functions of the transformed $T_i$ are independent of $\alpha$, in particular they coincide with the correlation functions of $T_i$ computed with the original field variables, which correspond to the choice $\alpha=0$.

Moreover, we can identify the physical operator subspace as composed from the field operators $\mathcal{F}$ in the cohomology of the BRST operator $\delta$, that is, $\delta\mathcal{F}=0$, $\mathcal{F}\neq \delta(\ldots)$ and $\mathcal{F}$ carries ghost number zero. Then clearly we can see that $\hat{\phi}$, $c$ form a doublet, as well as $\bar{c}$, $\chi$ and $\alpha$, $\beta$. Therefore by the doublet theorem \cite{Piguet:1995er}, the only non-trivial field in the physical subspace is $\phi$, proving that both theories are equivalent and formulated in terms of the physical field $\phi$.

\subsection{Implications}
The true power of the Equivalence Theorem was already realized by \cite{gomisweinberg,eqtheorem_blasi}, providing a way to give meaning to a nonrenormalizable theory. Indeed, starting from the theory in terms of $\hat{\phi}$ may lead one to believe that the obtained theory is nonrenormalizable based on standard powercounting arguments. However, as illustrated in the above subsection, the correlation functions of the physical field $\phi$ cannot change, meaning that $\phi=\hat{\phi}+\alpha \hat{\phi}^2 g(\alpha,\phi)$ is still the physical field. This simply means that the nonrenormalizable theory is written in terms of the ``wrong'' field variables and performing a change of variables can yield a perfectly renormalizable, physical theory.

The implications of the Equivalence Theorem are therefore limitless, as illustrated in for example \cite{eqtheorem_blasi}: one may always try to look for a renormalizable composite operator in terms of which to write the theory. Moreover, the Equivalence Theorem may be used as a tool to prove renormalizability, indeed when it would be easier to prove renormalizability in one particular setup, the Equivalence Theorem ensures renormalizability in all equivalent formulations. The downside of performing a change of variables is the fact that one may obtain an infinite number of interaction vertices, however the renormalization is perfectly under control as all coupling constants are related in terms of a finite number of parameters and the gauge parameter. This is of relevance for the predictability of the model of course, one does not want an infinite number of independent parameters.

In short, if an operator $F(\phi,\partial_\mu \phi,\dots)$ is renormalizable in the original field formulation, the transformed operator $F(\phi[\hat\phi],\partial_\mu \phi[\hat\phi],\dots)$ will also be in the transformed theory, and vice versa. 

\subsection{The Equivalence Theorem and admissible composite operators as new elementary fields}
In what follows, we will further enrich the Equivalence Theorem. For simplicity, let us explain in terms of a scalar field again. Assume $R(\phi)$ is a composite operator of the following type
\begin{equation}\label{n1}
R(\phi) = a\phi + \mathcal{O}(\phi^2)\end{equation}
where $\mathcal{O}(\phi^2)$ collect the higher order terms, $a$ can be any number. Note that $R(\phi)$ and $\phi$ carry the same quantum numbers, so the states they create from the vacuum will have non-vanishing overlap.

Evidently, we can immediately infer a related field transformation
\[\phi\longrightarrow R(\phi,\alpha):= a\phi + \alpha\mathcal{O}(\phi^2)\]
We can apply the Equivalence Theorem with this transformation, for any value of $\alpha$. 

Assume now that we want to use $R(\phi,\alpha)$ as our new elementary field variable, which necessitates that the correlation functions of $R(\phi,\alpha)$ must be finite (renormalizable) ones. Clearly, per Equivalence Theorem, the only admissible $R(\phi,\alpha)$ will be those that for specific value(s) of $\alpha$ were renormalizable operators in the original field variables.  Of course, in practice this will amount to first looking for renormalizable composite operators of the type \eqref{n1} in the original field theory, followed by a field transformation to these operators. Typically, $\alpha=1$ will be the only admissible choice for the parameter $\alpha$, as most likely for other values $R(\phi,\alpha)$ will not be renormalizable and thus not lead to a sensible new action with finite correlation functions for generic $R(\phi,\alpha)$.

This strategy is exactly what we will explicitly implement in the following for a gauge-Higgs theory. This adds the extra complication that the physical operators need to be gauge (BRST) invariant under the original BRST variation $s$ of the underlying gauge theory. As discussed in \cite{Dudal2021-md}, one can use the still nilpotent, combined BRST operator $\delta+s$ to control both gauge invariance and dependence on the new parameter $\alpha$, and thus the Equivalence Theorem in a gauge theory context. This boils down to considering a constrained cohomology of $s$ and $\delta$, see also \cite{Delduc:1996yh}.

\section{A gauge-invariant Abelian Higgs action}\label{beh}
\subsection{Short survey of some basic Higgs physics}
The aim of this section is to briefly introduce the Higgs field in our conventions, and as is done in many textbooks, so that we can also briefly highlight the issues with this approach, more specifically the fact that the elementary field excitations are not gauge-invariant and therefore unphysical.   

Following the FMS approach \cite{fmspaper,fmspaper2,fmsreviewmaas}, see also \cite{Maas:2017xzh,Maas:2023nsa,Sondenheimer:2019idq,Maas:2020kda,Dobson:2022ngz} for more or \cite{Binosi:2022ycu,Quadri:2023jcn,Quadri:2024aqo} for other works on gauge-invariant Higgs model extensions, we will introduce gauge-invariant composite operators, which we will interpret as physical particles and therefore observable in scattering experiments. Afterwards, we will perform a change of variables to obtain a theory where these FMS composite operators will be the new elementary (gauge-invariant) field excitations. We will make extensive use of the Equivalence Theorem, to prove equivalence of both setups, as well as renormalizability. Finally, we will argue on why we may remove the (transformed) gauge fixing term to obtain a new ``ungauged'' Higgs theory, as well as why we can also ignore the Jacobians arising from the non-linear pieces of the involved transformations.

The simplest realization of the Higgs mechanism \cite{broutenglertpaper,higgspaper,Higgs:1964ia,Higgs:1966ev,Guralnik:1964eu} is the Abelian Higgs model, describing a complex-valued scalar field $\phi$, coupled to a vector (gauge) field $A_\mu$ in the following way:
\begin{equation}
	S_{\text{Higgs}}=\int \dd[4]{x} \qty[\frac{1}{4} F_{\mu\nu}F_{\mu\nu} +(D_\mu \phi)^*D_\mu\phi+ \frac{1}{2} \lambda \qty(\phi^\dagger \phi -\frac{v^2}{2})].
\end{equation}
The field strength and covariant derivative are given by:
\begin{equation*}
	F_{\mu\nu}=\partial_\mu A_\nu-\partial_\nu A_\mu \,,\quad D_\mu =\partial_\mu +i e A_\mu.
\end{equation*}
Crucially, the Abelian Higgs model remains invariant under a gauge symmetry parameterized by the local gauge variation $\zeta(x)$,
\begin{equation}
	\phi(x)\rightarrow e^{ie\zeta(x)}\phi(x)\,,\quad A_\mu(x)\rightarrow A_\mu(x)-\partial_\mu \zeta(x).
\end{equation}
The classical potential is minimized for constant fields $\phi$, for which, $\abs{\phi}=\frac{v}{\sqrt{2}}$, therefore we will expand the complex field around for example $\phi=\frac{v}{\sqrt{2}}$ by splitting it into real and imaginary parts: \begin{equation}\phi=\frac{1}{\sqrt{2}}(v+h+i\rho).\end{equation} Then
\begin{equation}\label{eq:start-action}
    \begin{aligned}
    	S_\text{Higgs} &= \int \dd[4]{x} \Big[\frac{1}{4} F_{\mu\nu}F_{\mu\nu}+\frac{1}{2}(\partial_\mu h)^2+\frac{1}{2}(\partial_\mu \rho)^2 +\underbrace{\frac{e^2v^2}{2}A_\mu A_\mu}_\text{Mass term}+ ev A_\mu \partial_\mu \rho\\
    	& +\underbrace{\frac{1}{2}\lambda v^2 h^2}_\text{Mass term} - eA_\mu \rho\partial_\mu h +e A_\mu h \partial_\mu \rho+e^2vh A_\mu A_\mu +\frac{1}{2}e^2 \rho^2A_\mu A_\mu +\frac{1}{2}e^2 h^2 A_\mu A_\mu\\
    	& +\frac{1}{8}\lambda h^4+\frac{1}{8}\lambda \rho^4+\frac{1}{2}\lambda v h^3 + \frac{1}{2}\lambda v h\rho^2 + \frac{1}{4}\lambda h^2\rho^2\Big].
    \end{aligned}
\end{equation}
We can clearly see that this setup has a massive mode for $h$ with mass, $m^2=\lambda v^2$, a massless mode for $\rho$ while $A_\mu$ acquires a  mass of $e^2v^2$. The massless Goldstone boson is a consequence of the Goldstone theorem due to the gauge symmetry \cite{goldstonetheorem}.

The gauge symmetry transformations become
\begin{equation}
	A_\mu \rightarrow A_\mu -\partial_\mu \zeta \qc h\rightarrow h -e \zeta \rho \qc \rho \rightarrow \rho+e \zeta(v+h)
\end{equation}
By Elitzur's theorem \cite{elitzurtheorem} only gauge-invariant quantities can have a non-vanishing vacuum expectation value. Therefore, in order to interpret $h$, $A_\mu$ and $\rho$ as excitations of the vacuum, we need to fix the gauge. This is achieved by addition of a Faddeev-Popov gauge fixing term and introducing the necessary ghost fields $c$, $\bar{c}$ and multiplier $b$, where we will use the Landau gauge without loss of generality here. More precisely,
\begin{equation}
    S_\text{gf}=\int \dd[4]{x} \qty[ ib \partial_\mu A_\mu +\bar{c}\partial^2c ].
\end{equation}
The renormalizability of the Abelian Higgs model was discussed in \cite{Becchi:1974md}, with more details and subtleties discussed in \cite{Kraus:1995jk,Haussling:1996rq,Dudal2021-md}.

We will use the most general counterterms needed for renormalization of both $h$, $A_\mu$ and later on of the FMS operators as discussed in \cite{Dudal2021-md,Capri_2020}. It was also verified in \cite{Dudal2021-md}, that the configuration $\phi =\frac{v}{\sqrt{2}}$ becomes a true vacuum configuration by verifying the minimization of the vacuum energy (quantum effective action) as well as the vanishing of the tadpole diagrams for $\braket{h}$.

\subsection{Need for a gauge-invariant setup}
The goal of any theory is the computation of physical observables, that is, quantities that one can measure experimentally. A quantity depending on the gauge, might obtain non-zero expectation values in a gauge fixed setup, however that does not mean it is observable in nature. Of course, it also does not exclude physical information from being hidden in the gauge variant quantities, with as typical example the pole mass, which can be shown to be gauge parameter independent thanks to the Nielsen identities \cite{Nielsen:1975fs,Piguet:1984js} even if computed using a gauge variant correlation function.

Although that $h$ and $A_\mu$ can be interpreted as field excitations after fixing the gauge, however they are still not BRST invariant in general. Therefore these fields are neither physical nor can they be measured. A nice illustration of the problems of working with the gauge variant field variables can be found in e.g.~\cite{Dudal:2019aew} where it was shown that the K\"allén-Lehmann spectral representation \cite{peskin} of the elementary scalar field (the Higgs mode) is not gauge-invariant, and even shows non-positive behaviour of the spectral density, in sharp contrast with it being interpreted as a physical mode.

Only gauge-invariant operators can be physical and in principle, the spectrum of the theory, next to all measurable quantities, should be emerge from using those operators, this was already realized in \cite{fmspaper,fmspaper2}. However, generally speaking, this is not that surprising, as when using lattice gauge theory, gauge fixing is not necessary and all observables are to be computed using gauge-invariant operators, as all other expectation values would be zero \cite{elitzurtheorem}.

As explained in the thorough review  \cite{fmsreviewmaas}, we would like to interpret these gauge-invariant operators as corresponding to particles, therefore they must be both gauge-invariant and local. This requires them to be composite operators and in the Abelian Higgs model they are concretely given by
\begin{equation}\label{gaugeinvariantoperators}
    \begin{aligned}
    	O(x)&= \phi^\dagger\phi -\frac{v^2}{2}=\frac{1}{2}(h^2+\underline{2vh} +\rho^2) \\
    	V_\mu(x)&=-i\phi^\dagger D_\mu \phi=\frac{1}{2}\qty(-\rho \partial_\mu h+h\partial_\mu \rho+\underline{v\partial_\mu \rho}+\underline{eA_\mu(v^2}+h^2+2vh+\rho^2)).
    \end{aligned}
\end{equation}
These gauge-invariant operators were already scrutinized in \cite{Capri_2020,Dudal2021-md,Dudal:2019pyg,Dudal:2019aew}, including their renormalization properties. $O$ corresponds to the physical Higgs particles, while $V_\mu$ to a massive vector boson. For practical calculations this setup is perfectly sufficient, that is, treating these operators as composite operators of the original (elementary) field variables and compute with those in a certain gauge fixed setting.

However, we can and will perform a change of variables to obtain an equivalent theory in which the new elementary field excitations are those of $O$ and $V_\mu$ and to get rid of all remaining unphysical fields. Note that in \cite{Dudal:2019pyg} it was already explicitly shown that the FMS two-point correlation functions exhibit a physical K\"allén-Lehmann representation, including positive spectral density, indicative of unitarity.

The needed transformation comes from inverting \eqref{gaugeinvariantoperators}. Moreover, they are of the correct form such that the Equivalence Theorem holds, apart from a linear change of variables as can be seen from the underlined terms. Therefore both setups will dictate the same correlation functions and physics. Furthermore, the Equivalence Theorem proves renormalizability of the new model given the fact that these operators are renormalizable in the Abelian Higgs model \cite{Dudal2021-md,Capri_2020}. For the record, both operators are subject to mixing at the quantum level, $O$ with $v^2$ and $V_\mu$ with $\partial_\nu F_{\mu\nu}$. This will automatically be taken care of later on by transforming the most general counterterm---as constructed in \cite{Dudal2021-md} based on several constraining Ward identities--to the new variables. 

\subsection{Performing the change of variables}\label{polarparametrization}
\subsubsection{Intermezzo: the non-admissible gauge-invariant polar parametrization}
Following \cite{Dudal2021-md}, we first perform an intermediate change of variables to the polar parametrization of the Abelian Higgs model,
\begin{align}
	&\phi=\frac{1}{\sqrt{2}}(v+h')e^{i\rho'}.
\end{align}
The change of variables is defined as
\begin{equation}
    \begin{aligned}
    	&h= (h'+v)\cos\rho'-v \\
    	&\rho= (h'+v)\sin \rho' \\
    	&A_\mu = A_\mu'-\frac{1}{e}\partial_\mu \rho'
    \end{aligned}\qq{.}
\end{equation}
The Jacobian corresponding to this transformation may be written as follows
\begin{equation}
	\det \mqty(
		\cos\rho' & -(h'+v)\sin \rho' & 0 \\
		\sin \rho'& (h'+v)\cos\rho' & 0 \\
		0&-\frac{1}{e} \partial_\mu & \delta_{\nu\mu}
	) =\det\qty(h'+v)
\end{equation}
and exponentiated into the action using new scalar ghost fields $(\bar{\eta},\eta)$,
\begin{equation}\label{spook1}
		S_{\text{ghosts},1}=\int \dd[4]{x} ~\bar{\eta}(h'+v)\eta.
\end{equation}
Subsequently, in this polar parametrization, the BRST invariant observables $O$ and $V_\mu$ are expressed in terms of $h'$ and $A'_\mu$ as follows:
\begin{align}
	O=\frac{1}{2}(h'+v)^2-\frac{v^2}{2}\,,\quad V_\mu=\frac{e}{2}(h'+v)^2 A'_\mu.
\end{align}
After this change of variables, the classical action \eqref{eq:start-action} in the polar parametrization is given by the following expression:
\begin{equation}\label{polaraction}
    \begin{aligned}
    	S'_\text{Higgs}=&\int \dd[4]{x} \bigg[\frac{1}{4}   F^2(A'_\mu)+\frac{1}{2}(\partial_\mu h')^2+\frac{e^2}{2}(A')^2(v+h')^2+\frac{\lambda}{8}(h'^2+2h'v)^2 \\
    	&+ib \qty(\partial_\mu A'_\mu-\frac{1}{e}\partial^2\rho')+\bar{c}\partial^2c +\bar{\eta}(h'+v)\eta + J O +\Omega_\mu V_\mu\bigg]
    \end{aligned}
\end{equation}
where we now have also included external sources $J$ and $\Omega_\mu$ to couple the composite operators $O$ and $V_\mu$ to the action. The latter is still gauge fixed, therefore the action of the BRST symmetry generator $s$ becomes in the new parameterization
\begin{equation}
	sh'=0\,,\quad sA_\mu'=0\,,\quad s\rho'=ec\,,\quad sc=0\,,\quad s\bar{c}=ib\,,\quad sb=0\,,\quad s\bar{\eta}=s\eta=0.
\end{equation}
We see that $h'$ and $A'_\mu$ are already BRST invariant, so one might wonder why not stopping here and use these as new basic field variables?
This would however lead to problems, as neither of the two operators $h'$ and $A'_\mu$ are renormalizable in the original field formulation, and this will thus remain to be the case, which also follows from the Equivalence Theorem.
The nonrenormalizability can be easily verified by explicit calculation of the $h'$ and $A'_\mu$ two-point function up to one-loop.

\subsubsection{Decoupling of the polar ghosts from physical correlation functions}\label{ghosdec1}
We will now investigate the contributions from the ghost vertices to our theory, starting with the propagator as arising from \eqref{spook1}, yielding the constant propagator
\begin{equation}
	\Delta_{\bar{\eta}\eta}=\frac{1}{v},
\end{equation}
next to a single interaction vertex, $\int \dd[4]{x}~\bar{\eta}\eta h'$.

We will give the actual argument for ghost vertices of even a somewhat more general form than needed, namely we consider $F(h',A')$ a monomial in the fields $h'$ and $A'$, with $f_1$ certain constants,
\begin{equation}\label{ghostvertexform}
	\mathcal{S}_\text{ghosts,int}=\int \dd[4]{x} \bar{\eta}\eta F(h',A')=\int \dd[4]{x} \bar{\eta}\eta f_1\cdot  \frac{(h')^{m_1}}{m_1!} \frac{(A')^{m_2}}{m_2!}.
\end{equation}
Clearly, each such vertex has a Feynman rule of $-f_1$. We will now show that for correlation functions between physical operators ($O$, $V_\mu$ or in fact, any operator belonging to the BRST cohomology with ghost number zero), the ghost diagrams, from a theory with vertices such as \eqref{ghostvertexform}, will never contribute. We observe that at any such vertex, exactly $2$ ghost lines are created (or annihilated). We also observe that for each diagram between physical operators, there are no external ghost lines. Therefore, each ghost field we create by a ghost vertex, must be annihilated at another ghost vertex in the same diagram.
\begin{figure}
	\centering
	\begin{tikzpicture}
		\begin{feynman}
			\vertex (a1);
			\vertex[left= of a1](a2){\(\bar{\eta}_1\)};
			\vertex[right= of a1](a3){\(\eta_1\)};
			\vertex[above= of a1](a4){\(F_1\)};
			
			\vertex[right=6cm of a1] (b1);
			\vertex[left= of b1](b2){\(\bar{\eta}_2\)};
			\vertex[right= of b1](b3){\(\eta_2\)};
			\vertex[above= of b1](b4){\(F_2\)};
			
			\vertex[below=3cm of a1](c1);
			\vertex[above= of c1](c2){\(F_1\)};
			
			\vertex[right=6cm of c1] (d1);
			\vertex[left= of d1](d2){\(\bar{\eta}_2\)};
			\vertex[right= of d1](d3){\(\eta_2\)};
			\vertex[above= of d1](d4){\(F_2\)};
			
			\vertex[below=4cm of c1](e1);
			\vertex[left= of e1](e2){\(\bar{\eta}_1\)};
			\vertex[above= of e1](e4){\(F_1\)};
			
			\vertex[right=6cm of e1] (f1);
			\vertex[right= of f1](f3){\(\eta_2\)};
			\vertex[above= of f1](f4){\(F_2\)};
			
			\diagram*{
				(a2)--[ghost,edge label'=\(q\)](a1)--[ghost,edge label'=\(q+q_1\)](a3),
				(a1)--[edge label'=\(q_1\)](a4),
				
				(b2)--[ghost,edge label'=\(r\)](b1)--[ghost,edge label'=\(r+q_2\)](b3),
				(b1)--[edge label'=\(q_2\)](b4),

				(c1)--[edge label'=\(0\)](c2),
				c1--[ghost,loop,in=-45,out=-135,edge label'=\(q\), min distance=2cm]c1,
				
				(d2)--[ghost,edge label'=\(r\)](d1)--[ghost,edge label'=\(r+q_2\)](d3),
				(d1)--[edge label'=\(q_2\)](d4),
				
				(e2)--[ghost,edge label'=\(q\)](e1)--[ghost,edge label'=\(q+q_1\)](f1)--[ghost,edge label'=\(q+q_1+q_2\)](f3),
				(e1)--[edge label'=\(q_1\)](e4),

				(f1)--[edge label'=\(q_2\)](f4),
			};
		\end{feynman}
	\end{tikzpicture}
	
	\caption{Illustration on how to close or extend the ghost loop (dotted lines), for vertices where two ghost fields are created or annihilated.}\label{howtocloseghostloop}
\end{figure}
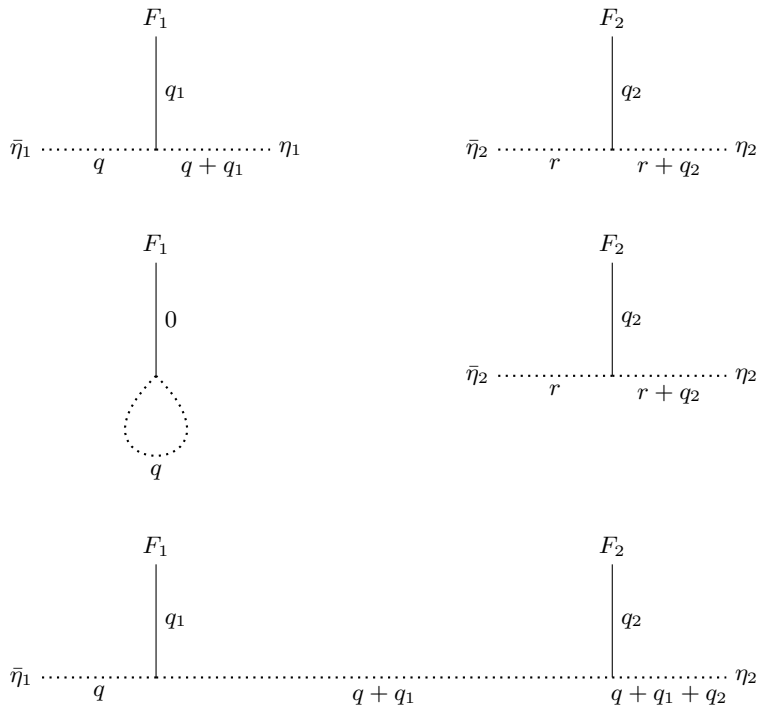
Suppose we calculate the contributions from a connected diagram, containing some ghost vertex, as shown in Figure \ref{howtocloseghostloop}, we must definitely annihilate $\eta_1$ as all internal lines must be closed. There are two choices, we can either choose to annihilate $\eta_1$ with $\bar{\eta}_1$ or we can annihilate with $\bar{\eta}_2$, corresponding to some other vertex with ghost fields. In the first case, we immediately find a closed ghost loop, while for the second case, we face the same question for $\eta_2$, and we can either keep extending the ghost line, or close the loop with $\bar{\eta}_1$. Given the fact that there are only a finite number of vertices in each diagram, at some point we must annihilate the ghost field $\eta_n$ with $\bar{\eta}_1$ and close the ghost loop, as there are no external ghost lines. Therefore for loop contributions to correlation functions of physical operators, containing ghost lines, there will always be at least one loop, consisting entirely of ghost propagators. Moreover, after applying momentum conservation at each vertex, we will first of all have that $\sum q_i=0$, but more importantly we recover the undetermined loop momentum $q$, following the notation of the Figure \ref{howtocloseghostloop}. The integration over this momentum will amount to the integration over a constant $C$, since the ghost propagator is a constant, moreover the ghost vertices will have as Feynman rules $-f_i$, also constants, such that we can write the contribution of each diagram with ghosts as, where the dots represent the other Feynman rules, propagators and integrations over other loop momenta:
\begin{equation*}
	D=\dots \underbrace{\int \frac{d^dq}{(2\pi)^d} C}_{\text{$=0$}}.
\end{equation*}
Crucially, the underbraced term becomes exactly zero in dimensional regularization, therefore the contributions from these ghost diagrams is always zero and they will never contribute to physics.

Furthermore, we stress that this argument holds up to all loop orders. Even though multi-loop contributions will seem to change the ghost two-point function, there will always be a closed loop of ghost propagators.
\subsubsection{At last: admissible transformation to elementary FMS operators }\label{fulltransfo}
In \ref{polarparametrization}, we have already formulated the Abelian Higgs model in a BRST invariant fashion, making use of the polar parametrization. To come to a renormalizable theory in terms of gauge-invariant operators, we will now formulate the theory in terms of the renormalizable, composite FMS-operators $O$ and $V_\mu$, by the following non-linear change of variables.
\begin{align}
	O&=\frac{1}{2}(h'+v)^2-\frac{v^2}{2} \,,\quad V_\mu=\frac{e}{2}(h'+v)^2A'_\mu,
\end{align}
or, inversely,
\begin{align}\label{laatstetransfo}
		h'&=-v+v\sqrt{1+\frac{2O}{v^2}}\,,\quad
	A'_\mu=\frac{2V_\mu}{ev^2}\frac{1}{1+\frac{2O}{v^2}}.
\end{align}
We will use these expressions as a field transformation of $h'$ and $A'_\mu$. First of all, we can write $h'$ and $A'_\mu$ as power series in $O$ and $V_\mu$ :
\begin{equation}\label{trans2}
    \begin{aligned}
	h' &= \frac{O}{v} - \frac{O^2}{2v^3} + \frac{O^3}{2v^5} + \ldots \\
    	A' &= \frac{2V_\mu}{ev^2} - \frac{2V_\mu}{ev^2} \frac{2O}{v^2} + \frac{2V_\mu}{ev^2} \frac{4O^2}{v^4} + \ldots
    \end{aligned}\qq{.}
\end{equation}
After applying this transformation, the classical Lagrangian (without the counterterms) transforms as follows and we omitted here the ghost contributions related to the Jacobians (see above and below in the manuscript).
\begin{equation}\label{eq:final-action}
    \begin{aligned}
    	\mathcal{S}'&=\int \dd[4]{x}\Bigg[\frac{2}{e^2v^4}\frac{1}{(1+\frac{2O}{v^2})^2}\qty(V_\mu (-\partial^2)V_\mu + V_\mu \partial_\mu \partial_\nu V_\nu ) \\
        & + \frac{16}{e^2v^8}\frac{1}{(1+\frac{2O}{v^2})^4}\Big(\frac{V^2}{2}(\partial_\mu O)^2-\frac{V_\mu V_\nu}{2}\partial_\mu O \partial_\nu O\Big) \\
    	& + \frac{(\partial_\mu O)^2}{2v^2(1 + \frac{2O}{v^2})} + \frac{2}{v^2} \frac{V^2}{1+\frac{2O}{v^2}} + \frac{\lambda}{2} O^2 + ib \qty[\frac{2}{ev^2} \partial_\mu\qty(\frac{V_\mu}{1+\frac{2O}{v^2}}) - \frac{1}{e} \partial^2\rho' ] + \bar{c}\partial^2 c\Bigg].
    \end{aligned}
\end{equation}
As this is simply a redefinition of the variables used to express the theory, the most general counterterm should be taken over from the original theory, which was derived in \cite{Dudal2021-md}.
In the new variables the counterterms are given by
\begin{equation}
\begin{aligned}\label{gaugeinv_ct}
	\delta \mathcal{S}'&= \int \dd[4]{x}\Bigg[a_0\Bigg(\Big(\frac{2}{ev^2}\Big)^2\frac{1}{(1+\frac{2O}{v^2})^2}\frac{1}{2}\Big(V_\mu (-\partial^2)V_\mu + V_\mu \partial_\mu \partial_\nu V_\nu\Big)\\
    &+ \frac{16}{e^2v^8}\frac{1}{(1+\frac{2O}{v^2})^4}\Big(\frac{V^2}{2}(\partial_\mu O)^2-\frac{V_\mu V_\nu}{2}\partial_\mu O \partial_\nu O\Big)\\
	&+\frac{\Big(-\partial^2\delta_{\mu\nu}+\partial_\mu \partial_\nu\Big)\Omega_\nu}{e^2v^2}\frac{V_\nu}{1+\frac{2O}{v^2}}+\frac{1}{8e^2}\Big(\Omega_\mu (-\partial^2)\Omega_\mu +\Omega_\mu \partial_\mu \partial_\nu \Omega_\nu\Big)\Bigg)\\
	&+a_1\Bigg(\frac{(\partial_\mu O)^2}{2v^2(1+\frac{2O}{v^2})}+\frac{2}{v^2}\frac{V^2}{1+\frac{2O}{v^2}}+\Omega_\mu V_\mu +\frac{1}{8}v^2\Omega_\mu \Omega_\mu +\frac{1}{4}O\Omega_\mu\Omega_\mu\Bigg)\\
	&+a_2\Bigg(\frac{\lambda}{2}O^2+JO-\frac{1}{4}O\Omega_\mu \Omega_\mu+\frac{1}{32\lambda}\Big(\Omega_\mu \Omega_\mu \Omega_\nu \Omega_\nu+16J^2-8J\Omega_\mu \Omega_\mu\Big)\Bigg)\\
	&+\delta a\Bigg(\frac{v^4}{4}+\frac{1}{16\lambda^2}\Big(\Omega_\mu \Omega_\mu \Omega_\nu \Omega_\nu+16J^2-8J\Omega_\mu \Omega_\mu\Big)-\frac{1}{\lambda}\Big(Jv^2-\frac{1}{4}v^2\Omega_\mu \Omega_\mu\Big)\Bigg)\\
	&+\delta \sigma\Bigg(v^2O+\frac{1}{\lambda}\Big(Jv^2-\frac{1}{4}v^2\Omega_\mu \Omega_\mu-2JO+\frac{1}{2}O\Omega_\mu \Omega_\mu\Big)\\&-\frac{1}{8\lambda^2}\Big(\Omega_\mu \Omega_\mu \Omega_\nu \Omega_\nu+16J^2-8J\Omega_\mu \Omega_\mu\Big)\Bigg)\Bigg]
\end{aligned}
\end{equation}
where we can reuse the 1-loop results calculated with \((h,A_\mu)\) as variables \cite{Capri_2020,Dudal2021-md}.
There relevant counterterm parameters are given by
\begin{equation}\label{eq:counterterm}
    \begin{aligned}
        (\delta \sigma)^{\text{1-loop}}&=-\frac{1}{(4\pi)^{d/2}}\frac{1}{v^2}\qty(\frac{3}{2} \lambda \Gamma(1-d/2) (m_h^2)^{d/2-1}+e^2(d-1) \Gamma(1-d/2) (m_A^2)^{d/2-1}),\\
    	(\delta a)^{\text{1-loop}}&=-\frac{1}{v^2} \frac{\Gamma(1-d/2)}{(4\pi)^{d/2}}\qty(e^2(d-1)(m_A^2)^{d/2-1}+\lambda (m_h^2)^{d/2-1}), \\
        (a_1)^{\text{1-loop}} &= \frac{3e^2}{(4\pi)^2}\qty(\frac{2}{\varepsilon}-\gamma + \ln(4\pi)), \\
        (a_2)^{\text{1-loop}} &= \frac{5\lambda - 6 \frac{e^4}{\lambda}}{(4\pi)^2} \qty(\frac{2}{\varepsilon}-\gamma + \ln(4\pi)),
    \end{aligned}
\end{equation}
these will be used later on for the \(O\)-propagator, and crucially do not require modification for the gauge invariant setup.
\(\delta\sigma\) and \(\delta a\) are related to the vanishing of tadpoles and the minimization of the quantum effective action and were explicitly determined in \cite{Dudal2021-md}.
Notice that we left the original sources present, that is, $J$ coupled to $O$ and $\Omega_\mu$ to $V_\mu$.
Derivatives of the generating functional w.r.t.~these sources will define the (renormalized) connected Green's functions of these operators, including the already alluded to mixing at the quantum level.

Finally, the relevant vertices of the complete, renormalized action are given in the Appendix \ref{apprules}.

Both field transformations are of the form $\phi=\hat{\phi} +\hat{\phi}^2 g(\phi)$, apart from a linear rescaling, consequently the Equivalence Theorem holds. Therefore $O$ and $V_{\mu}$ remain renormalizable in the transformed theory, as this was the case in the Abelian Higgs model. 
Crucially, the elementary field excitations of the gauge-invariant theory are now exactly $O$ and $V_\mu$, by construction, and their correlation functions will be finite, despite the complicated nature of the action \eqref{eq:final-action}. 

Moreover, the BRST transformation is as follows:
\begin{equation}\label{brstbijna}
	sO=0\,,\quad sV_\mu=0\,,\quad s\rho'=ec\,,\quad sc=0\,,\quad s\bar{c}=ib\,,\quad sb=0\,,\quad s\bar{\eta}=s\eta=0.
\end{equation}
Therefore the physical subspace is once more built solely from $O$ and $V_\mu$, as the other fields appear in doublets. This makes explicit the fact that any physical, gauge-invariant observable may be built from $O$ and $V_\mu$.

As an important side comment, we have verified in leading order perturbation theory  that the quantum effective action remains minimized in terms of the parameter $v$ in the Appendix \ref{appvac}, as well as that tadpole diagrams potentially contributing to $\expval{O}$ add up to zero in the Appendix \ref{apptad}. 

\subsubsection{More decoupling ghost fields}\label{ghostvanishagain}
As before, there will be new ghosts arising from the Jacobian determinant from the non-linear field transformation \eqref{trans2}. At the level of the action, this becomes via the introduction of extra ghost fields $(\bar{\omega}, \omega)$ and $(\omega_\mu,\bar{\omega}_\mu)$
\begin{equation}\label{ghostsfulltransfo}
	S_{\text{ghosts}, 2}=\int \dd[4]{x}
	\begin{bmatrix}
		\bar{\omega}&\bar{\omega}_\mu
	\end{bmatrix}
	\begin{bmatrix}
		\frac{1}{v}\frac{1}{\sqrt{1+\frac{2O}{v^2}}}& 0 \\
		-\frac{4}{ev^4}V_\mu \frac{1}{(1+\frac{2O}{v^2})^2}& \delta_{\mu\nu}\frac{2 }{ev^2}\frac{1}{1+\frac{2O}{v^2}}
	\end{bmatrix}
	\begin{bmatrix}
		\omega\\
		\omega_\nu	
	\end{bmatrix}.
\end{equation}
Now immediately, we obtain the constant ghost propagators
\begin{equation}
	\Delta_{\bar{\omega}\omega}=v \,,\quad \Delta_{\bar{\omega}_\mu \omega_\nu}= \frac{ev^2}{2}\delta_{\mu\nu}.
\end{equation}
Expanding the square root and the fractions in \ref{ghostsfulltransfo}, we can see that all ghost terms are again of the form:
\begin{equation*}
	\int \dd[4]{x}~\bar{\omega}\omega F(O,V).
\end{equation*}
Therefore, correlation functions between the physical excitations, $O$ and $V$, will always create exactly two ghosts of the new type. Subsequently, we may repeat the reasoning in \ref{ghosdec1}, to obtain that this second ghost sector will also not contribute to any relevant physics and can thus be safely ignored.

\subsubsection{Decoupling of the remnants of the gauge fixing}
In the previous sections, we have obtained a theory written in terms of the excitations $O$, $V_\mu$ and $\rho'$, $b$, $\bar{c}$, $c$ and ghost fields corresponding to the changes of variables. We will now argue why we can remove that gauge fixing term from the final action \eqref{eq:final-action}, to obtain an ``ungauged'' Higgs model.

This is perhaps not unexpected: as the action is now written in terms of gauge-invariant variables that are path integrated over, there is no need anymore to gauge fix anything to eliminate the redundant degrees of freedom or to circumvent Elitzur's theorem \cite{elitzurtheorem}. To construct \eqref{eq:final-action}, we however had to start from the gauge fixed original Higgs model. The imposed gauge choice $G(A_\mu)=\partial_\mu A_\mu=0$ was enforced by addition of the following term the action, here already rewritten in terms of the new variables:
\begin{equation}\label{ijkterm}
    \int \dd[4]{x} \left[ib\Big[\frac{2}{ev^2}\partial_\mu\Big(\frac{V_\mu}{1+\frac{2O}{v^2}}\Big)-\frac{1}{e}\partial^2\rho'\Big]+\bar{c}\partial^2 c\right].
\end{equation}
First of all, the propagators in the gauge-invariant model are given by the following expressions.
\begin{equation}
\begin{aligned}
	&\Delta_{V_\mu V_\nu}=\frac{e^2 v^4}{4} \frac{\delta_{\mu\nu}+\frac{p_\mu p_\nu}{m^2}}{p^2+m_A^2} \hspace{0.5cm}\Delta_{O O}= \frac{v^2}{p^2+m_h^2} \hspace{0.5cm} \Delta_{\rho'\rho'}=\frac{1}{p^2 v^2}\\ 
    &\Delta_{V_\mu \rho'}= i\frac{p_\mu}{2p^2} \hspace{0.5cm}\Delta_{b\rho'}=-i\frac{e}{p^2} \hspace{0.5cm} \Delta_{\bar{c}c}=\Delta_{\bar{c}c}=-\frac{1}{p^2}.
\end{aligned}
\end{equation}
One may easily verify that the removal of the above gauge fixing terms will not influence the tree level-propagators of neither $O$ nor $V_\mu$.

Second of all, there are no interaction vertices creating or annihilating the $\rho'$ field. Therefore, when a Feynman diagram contributing to a physical correlation function would introduce a $bVO^n$ vertex, corresponding to the first term in the gauge fixing term, we must annihilate the $b$-field with another $b$-field, as a vertex creating a $\rho'$ field does not exist. But given that the $bb$-propagator is actually zero, these vertices cannot contribute to physical correlation functions. Similarly, the $ib \partial^2 \rho'$-term will also not contribute to physical correlation functions.

Moreover, the $\bar{c}\partial^2 c$-term fully decouples from the other fields, therefore we can also ignore this term for what concerns the physical correlation functions.

Summarizing, we have argued that each individual term term of the gauge fixing term does not contribute to physical correlation functions between $O$ and $V_\mu$. Said otherwise, we can omit \eqref{ijkterm} integrally.

\section{Explicit illustration: Propagator of the gauge-invariant scalar $O$}
We can clearly see that the $V$-propagator at tree level is a Proca-type propagator, commonly known to lead to a powercounting nonrenormalizable theory. It certainly is the kind of propagator one expects for a gauge-invariant massive vector particle.  

As we have just proven, including loop corrections, the $V$-propagator should remain renormalizable as a consequence of the Equivalence Theorem.  A priori,  this is highly non-trivial, as the longitudinal part of a Proca-propagator scales like a constant rather than displaying a fast enough fall-off in the UV. A first hint to why this can still work out is by inspecting the form of the counterterm vertices. For example, the $OVV$ vertex will yield a Feynman rule of:
\begin{align}
	\Gamma_{OVV}(p_1,p_2,p_3)&=\frac{8}{v^4}\Bigg(\Big(\frac{p_2^2}{m_A^2}P_{\mu\nu}(p_2) +\frac{p_3^2}{m_A^2}P_{\mu\nu}(p_3)\Big)(1+a_0)+\delta_{\mu\nu}(1+a_1)\Bigg)(2\pi)^d \delta(\sum_i p_i).
\end{align}
We see that this vertex factor contains the projection operator
\begin{equation}
    P_{\mu\nu}(p) = \eta_{\mu\nu} - \frac{p_\mu p_\nu}{p^2}\qc
\end{equation}
which will project the propagator unto its (well-behaved) transversal part in momentum space.

In this section we will explicitly show that the \(O\)-propagator is the same in the fully gauge-invariant setup as when using the base \(h,A_\mu\) fields variables \cite{Dudal:2019pyg}.
The \(O\)-propagator can be constructed as usual by taking functional derivatives with respect to the external source \(J\) of the path integral, but as \(J\) also couples directly (linearly and quadratically) to counterterms we are required to keep track of \(J\) in the related Feynman diagrams.
To continue, we need to calculate all one-loop diagrams with two external sources \(J\), which are given by

\noindent\makebox[\textwidth][c]{
\begin{tabular}{ccc}
    \parbox{5.5cm}{\begin{feyndiag}[baseline=(i)]
        \label{diag:trivial};
        \coordinate (i) at (-1cm,0);
        \draw[dotted] (-1.5cm,0) node[left] {\(J\)} -- (-1cm,0) (1cm,0) -- (1.5cm,0) node[right] {\(J^\dag\)};
        \draw[scalar] (-1cm,0) -- (1cm,0);
    \end{feyndiag}} & \sunset[5.5cm]{scalar}{photon}[photon] & \sunset[5.5cm]{scalar}{scalar} \\
    \snail[5.5cm]{scalar} & \snail[5.5cm]{scalar}[photon] & \parbox{5.5cm}{\begin{feyndiag}[baseline=(i)]
        \label{diag:tadpole};
        \coordinate (i) at (-1cm,0);
        \draw[dotted] (-1.5cm,0) node[left] {\(J\)} -- (-1cm,0) (1cm,0) -- (1.5cm,0) node[right] {\(J^\dag\)};
        \draw[scalar] (-1cm,0) -- (1cm,0);
        \draw[scalar] (0,0) -- (0,0.5cm);
        \filldraw[scalar,pattern=north east lines] (0,1cm) circle[radius=0.5cm];
    \end{feyndiag}} \\
    \multicolumn{3}{c}{\parbox{\textwidth}{\refstepcounter{feyndiag}\begin{equation}
        \label{diag:ct}
        \begin{tikzpicture}[baseline=(ct)]
            \node[crossed dot] (ct) at (0,0) {};
            \draw[dotted] (-1.5cm,0) node[left] {\(J\)} -- (-1cm,0) (1cm,0) -- (1.5cm,0) node[right] {\(J^\dag\)};
            \draw[scalar] (-1cm,0) -- (ct.west) (ct.east) -- (1cm,0);
        \end{tikzpicture} + 
        \begin{tikzpicture}[baseline=(ct)]
            \node[crossed dot] (ct) at (1cm,0) {};
            \draw[dotted] (-1cm,0) node[left] {\(J\)} -- (-0.5cm,0) (ct.east) -- (1.5cm,0) node[right] {\(J^\dag\)};
            \draw[scalar] (-0.5cm,0) -- (ct.west);
        \end{tikzpicture} +
        \begin{tikzpicture}[baseline=(ct)]
            \node[crossed dot] (ct) at (-1cm,0) {};
            \draw[dotted] (-1.5cm,0) node[left] {\(J\)} -- (ct.west) (0.5cm,0) -- (1cm,0) node[right] {\(J^\dag\)};
            \draw[scalar] (ct.east) -- (0.5cm,0);
        \end{tikzpicture} + 
        \begin{tikzpicture}[baseline=(ct)]
            \node[crossed dot] (ct) at (0,0) {};
            \draw[dotted] (-0.5cm,0) node[left] {\(J\)} -- (ct.west) (ct.east) -- (0.5cm,0) node[right] {\(J^\dag\)};
        \end{tikzpicture}
        \tag{CT}
    \end{equation}}}
\end{tabular}
}
where the dotted lines stand for the source \(J\), solid lines for the scalar \(O\) and wavy lines for \(V\).
Counterterms are denoted with crosses and the shaded loop is equal to the connected one-point function \(\expval{O}_c^\text{1-loop}\).
The explicit expressions for the diagram contributions are given in Appendix \ref{apx:explicit-diagrams}.
The one-loop two-point function can be written in the form
\begin{align*}
	\expval{O(p)O(-p)}=\frac{v^2}{p^2+m_h^2}+\frac{\Pi_{OO}(p^2)}{(p^2+m_h^2)^2}
\end{align*}
which, after summing all contributions, takes the form
\begin{equation}
\begin{aligned}
	\Pi_{OO}(p^2)&=\int_0^1dx\frac{\Gamma(2-d/2)}{(4\pi)^{d/2}}\Big(\frac{p^4}{2}-2p^2m_h^2+2m_h^4\Big)H(m_h^2,m_h^2,p^2)^{d/2-2}\\
	&+\frac{\Gamma(2-d/2)}{(4\pi)^{d/2}}(2(d-1)m_A^4+2p^2m_A^2+\frac{p^4}{2})H(m_A^2,m_A^2,p^2)^{d/2-2}\\
	&-\frac{\Gamma(1-d/2)}{(4\pi)^{d/2}}p^2(m_A^2)^{d/2-1}\\
	&+v^2m_h^2\Big(-2\frac{(\delta a)}{\lambda}\Big)+p^2v^2\Big(-a_1+4\frac{(\delta \sigma)}{\lambda}-4\frac{(\delta a)}{\lambda}\Big)+p^4\Big(-\frac{a_2}{\lambda}+4\frac{(\delta \sigma)}{\lambda^2}-2\frac{(\delta a)}{\lambda^2}\Big)
\end{aligned}
\end{equation}
with \(H(m_1^2,m_2^2,p^2)=x(1-x)p^2+(1-x)m_1^2 +x m_2^2\).
We can infer that this is exactly equal to the expression found in \cite[(2.32)]{Dudal:2019pyg} where this correlation function was computed in terms of the original (gauge variant field variables). In particular, the counterterms \eqref{eq:counterterm} nicely cancel all divergences in the above two-point function, despite the complicated nature of the underlying gauge-invariant
action that was used. Indeed, upon substituting the one-loop values for $a_1,a_2,\delta a$ and $\delta \sigma$ from \eqref{eq:counterterm}, we obtain for the would-be divergent piece:
\begin{equation}
\begin{aligned}
	\!\!\!\Pi_{OO}^{\text{div}}(p^2)=&\frac{\frac{2}{\epsilon}-\gamma+\log 4\pi}{(4\pi)^2}
    \qty[\qty(2m_h^4+6m_A^4-2v^2m_h^2(\lambda+\frac{3e^4}{\lambda}))+p^2\qty(-2m_h^2+3m_A^2-3m_A^2+2m_h^2)+p^4(1-1)] \\
	=&0.	
\end{aligned}
\end{equation}
As such, we end up with a finite expression
\begin{align}
	\Pi_{OO}(p^2)=&-\frac{1}{(4\pi)^2}\int_0^1dx\Bigg[\Big(\frac{p^4}{2}-2p^2m_h^2+2m_h^4\Big)\log \frac{p^2x(1-x)+m_h^2}{\mu^2}\nonumber\\
	&+\Big(6m_A^4+2p^2m_A^2+\frac{p^4}{2}\Big)\frac{p^2x(1-x)+m_A^2}{\mu^2} +4m_A^4-m_A^2 p^2 \Big(1-\log \frac{m_A^2}{\mu^2}\Big)\nonumber\\
	&+(2m_h^4-2p^2m_h^2-4p^4)\Big(1-\log \frac{m_h^2}{\mu^2}\Big)+\Big(2m_A^4-2p^4\frac{e^4}{\lambda^2}\Big)\Big(1-3\log \frac{m_A^2}{\mu^2}\Big)\Bigg].
\end{align}
This is an explicit illustration of the power of the Equivalence Theorem. Evidently, the pole mass of the gauge-invariant scalar will equal that already discussed in \cite{Dudal:2019pyg} upon using the same resummation scheme.

\section{Conclusion}
We have reconsidered the Equivalence Theorem in quantum (gauge) field theory algebraically, using an extended BRST symmetry. In general, physical observables and correlation functions do not depend on a local change of variables, allowing one to look at power-counting nonrenormalizable theories from a different perspective \cite{gomisweinberg,eqtheorem_blasi}. We introduced the subclass of new admissible field operators that we can transform our field variables into. More precisely, these are the gauge-invariant, composite operators of the original elementary fields, displaying the same quantum numbers, so overlapping states are created from the vacuum by these operators. In the context of gauge-Higgs theories, these operators were constructed by Fr\"ohlich--Morchio--Strocchi \cite{fmspaper,fmspaper2,fmsreviewmaas}.

The upshot of doing a field transformation to these FMS operators is that the eventual action is then expressed in terms of gauge-invariant operators, correctly capturing the physical spectrum, and all computations are manifestly gauge-invariant, order per order and even diagram per diagram when working perturbatively. We showed that several Jacobian-related contributions, including the remnants of the original gauge fixing, fully decouple from the physical correlation functions. The renormalizability is guaranteed as a consequence of the Equivalence Theorem. Our strategy is thus an explicit realization of the elsewhere dubbed ``augmented perturbation theory'', see e.g.~ \cite{Maas:2017xzh,Sondenheimer:2019idq,Maas:2020kda,Dobson:2022ngz}, in terms of the FMS operators, where the underlying computations still happen using the original fields.

As a proof of principle, we have applied the Equivalence Theorem to the Abelian Higgs model and its FMS operators and explicitly computed the one-loop gauge-invariant scalar Green's function and pole mass.

The next step in this research program would be to apply this framework to an $SU(2)$ gauge-Higgs theory with fundamental matter, followed by further generalizations to the $SU(2)_L\times U(1)$ sector of the Standard Model \cite{Dudal:2023jsu,Peruzzo:2024heb}, to render a fully gauge-invariant accounting of electroweak physics.

As in \cite{fmsreviewmaas}, we want to stress once more there that this transition from a description in terms of gauge variant elementary fields to gauge-invariant composite operator fields is not just a theoretical curiosity. Indeed, subtle physical differences can occur, albeit tiny and thus not yet quantitatively observed.
In particular because of the bound state nature of the composite description, e.g.\ see \cite{Maas:2024hcn,Jenny:2022atm,Dobson:2022ngz} for suggestions in this direction. 

One could also question the physical relevance of these and our current works based on the BRST invariance of the original theory itself. Indeed, the BRST invariance also leads to Nielsen identities \cite{Nielsen:1975fs,Piguet:1984js} which allow to prove e.g.~the gauge parameter independence of e.g.~pole masses or vacuum energies. Note that Nielsen identities do not allow to prove the gauge invariance of these aforementioned quantities though, only the less strong gauge parameter independence per class of gauges. In some gauges, due to infrared singularities, the consequences of the Nielsen identities can become less clear \cite{Aitchison:1983ns,Das:2013iha}. Such problems will not occur in an explicitly gauge-invariant setting. Another manifestation of the problems caused by using the gauge variant degrees of freedom is that their spectral functions are not even gauge parameter independent, including issues with the pole mass residues, see for instance \cite{Dudal:2019aew}. Overall, we hope that our main message is clear: the more gauge-invariant the methodology (insofar possible), the better to capture the correct physics in a gauge theory. Typical examples are QCD lattice gauge theory or, indeed, the FMS approach for gauge-Higgs systems.

\section*{Acknowledgments}
We thank G.~Peruzzo and S.P.~Sorella  for useful feedback. The work of D.~Dudal and T.~Oosthuyse was supported by KU Leuven IF project C14/21/087.

\appendix
\FloatBarrier
\renewcommand\thefigure{\thesection.\arabic{figure}}

\section{Some relevant Feynman rules in the gauge-invariant formulation}\label{apprules}
There are an infinite number of vertices arising from the action \eqref{eq:final-action}, but we will only show the vertices that play a role at one-loop for what concerns the $O$-propagator and effective action. Concretely:
\setcounter{figure}{0}
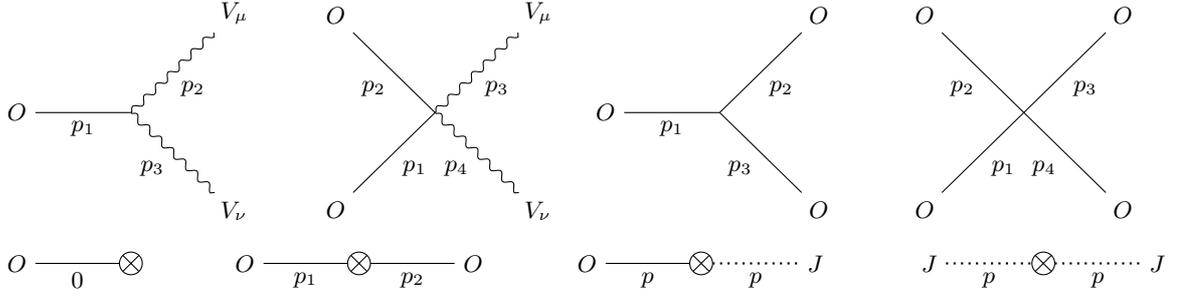
\begin{figure}
	\centering
	\begin{tikzpicture}
		\begin{feynman}
			\vertex (a1) {\(O\)};
			\vertex[right= of a1] (a2);
			\vertex[above right= of a2] (a3) {\(V_\mu\)};
			\vertex[below right= of a2] (a4) {\(V_\nu\)};
			
			\vertex[right=4cm of a2] (b2);
			\vertex[above left= of b2] (b1) {\(O\)};
			\vertex[below left= of b2] (b0) {\(O\)};
			\vertex[above right= of b2] (b3) {\(V_\mu\)};
			\vertex[below right= of b2] (b4) {\(V_\nu\)};
			
			\vertex[right=2cm of b2] (c1) {\(O\)};
			\vertex[right= of c1] (c2);
			\vertex[above right= of c2] (c3) {\(O\)};
			\vertex[below right= of c2] (c4) {\(O\)};
			
			\vertex[right=4cm of c2] (d2);
			\vertex[above left= of d2] (d1) {\(O\)};
			\vertex[below left= of d2] (d0) {\(O\)};
			\vertex[above right= of d2] (d3) {\(O\)};
			\vertex[below right= of d2] (d4) {\(O\)};

			\vertex[below=2cm of a1] (i1) {\(O\)};
			\vertex[crossed dot,right= of i1] (i2){};
			
			\vertex[right= of i2] (j1) {\(O\)};
			\vertex[crossed dot,right= of j1] (j2){};
			\vertex[right= of j2] (j3){\(O\)};
			
			\vertex[right= of j3] (k1) {\(O\)};
			\vertex[crossed dot,right= of k1] (k2){};
			\vertex[right= of k2](k3){\(J\)};
			
			\vertex[right= of k3] (l1) {\(J\)};
			\vertex[crossed dot,right= of l1](l2){};
			\vertex[right= of l2](l3){\(J\)};
			\diagram*{
				(a1)--[edge label'=\(p_1\)](a2)--[boson,edge label'=\(p_2\)](a3),
				(a2)--[boson,edge label'=\(p_3\)](a4),
				
				(b0)--[edge label'=\(p_1\)](b2)--[boson,edge label'=\(p_3\)](b3),
				(b1)--[edge label'=\(p_2\)](b2)--[boson,edge label'=\(p_4\)](b4),
				
				(c1)--[edge label'=\(p_1\)](c2)--[edge label'=\(p_2\)](c3),
				(c2)--[edge label'=\(p_3\)](c4),
				
				(d0)--[edge label'=\(p_1\)](d2)--[edge label'=\(p_3\)](d3),
				(d1)--[edge label'=\(p_2\)](d2)--[edge label'=\(p_4\)](d4),
				
				(i1)--[edge label'=\(0\)](i2),
				
				(j1)--[edge label'=\(p_1\)](j2)--[edge label'=\(p_2\)](j3),
				
				(k1)--[edge label'=\(p\)](k2)--[ghost,edge label'=\(p\)](k3),
				
				(l1)--[ghost,edge label'=\(p\)](l2)--[ghost,edge label'=\(p\)](l3),
			};
		\end{feynman}
	\end{tikzpicture}
	\caption{Vertices contributing to the one-and two-point function of $O$.}
    \
\end{figure}
\begin{align*}
	\Gamma_{OVV}(p_1,p_2,p_3)&=\frac{8}{v^4}\Bigg(\Big(\frac{p_2^2}{m_A^2}P_{\mu\nu}(p_2) +\frac{p_3^2}{m_A^2}P_{\mu\nu}(p_3)\Big)(1+a_0)+\delta_{\mu\nu}(1+a_1)\Bigg)(2\pi)^d \delta(\sum_i p_i),\\
	\Gamma_{OOVV}(p_1,p_2,p_3,p_4)&=-\frac{48}{v^6}\Bigg(\Big(\frac{p_3^2}{m_A^2}P_{\mu\nu}(p_3) +\frac{p_4^2}{m_A^2}P_{\mu\nu}(p_4)\Big)(1+a_0)+\frac{2}{3}\delta_{\mu\nu}(1+a_1)\Bigg)(2\pi)^d \delta(\sum_i p_i),\\
	&+\frac{16}{m_A^2v^6}\Bigg(2p_1\cdot p_2\delta_{\mu\nu}-p_{1,\mu}p_{2,\nu}-p_{2,\mu}p_{1,\nu}\Bigg)(1+a_0)(2\pi)^d \delta(\sum_i p_i),\\
	\Gamma_{OOO}(p_1,p_2,p_3)&=\frac{1}{v^4}(p_1^2+p_2^2+p_3^2)(1+a_2)(2\pi)^d \delta(\sum_i p_i),\\
	\Gamma_{OOOO}(p_1,p_2,p_3,p_4)&=-\frac{4}{v^6}(p_1^2+p_2^2+p_3^2+p_4^2)(1+a_2)(2\pi)^d \delta(\sum_i p_i),\\
	\Gamma_{bVO}(p_1,p_2,p_3)&=p_{1,\mu}\frac{4}{ev^4}(2\pi)^d \delta(\sum_i p_i),\\
	\Gamma_{bVOO}(p_1,p_2,p_3,p_4)&=-p_{1,\mu}\frac{16}{ev^6}(2\pi)^d \delta(\sum_i p_i).
\end{align*}
We will also list some counterterm vertices that were used, but note again that there are infinitely many more coming from the most general counterterm \eqref{gaugeinv_ct}:
\begin{align*}
	\Gamma_{OO}(p_1,p_2)&=(-a_2 \lambda - a_1 \frac{p_1^2}{v^2})(2\pi)^d \delta (p_1+p_2),\\
	\Gamma_{O}(p)&=-(\delta \sigma) v^2(2\pi)^d\delta(p),\\
	\Gamma_{JO}(p)&=-\Big(1+a_2-2\frac{\delta \sigma}{\lambda}\Big)\tilde{J}(-p),\\
	\Gamma_{JJ}&=\Big(-\frac{a_2}{\lambda}-2\frac{\delta a}{\lambda^2}+4\frac{\delta \sigma}{\lambda^2}\Big)\tilde{J}(p)\tilde{J}(-p).
\end{align*} 
 
\section{Minimization of the quantum effective action in the gauge-invariant formulation}\label{appvac}
Let us check whether the a priori classical parameter $v$ still minimizes the (one-loop) quantum effective action. Note that in the gauge-invariant formulation, at classical order, $v$ is a free (input) parameter of the theory. At one-loop, one gets
\begin{align*}
	\Gamma[v]&=\frac{d-1}{2}\int \frac{d^d q}{(2\pi)^d} \log \Big( p^2+e^2v^2\Big)+(\delta a) \frac{v^4}{4}+\frac{1}{2}\int \frac{d^dq}{(2\pi)^d} \log\Big(p^2  +\lambda v^2\Big),
\end{align*}
thus
\begin{align*}
	\frac{\delta \Gamma[v]}{\delta v}&= \int \frac{d^dq}{(2\pi)^d} \frac{e^2v(d-1) }{p^2+e^2v^2}+\int \frac{d^d q}{(2\pi)^d}\frac{\lambda v}{p^2+\lambda v^2}+(\delta a)v^3=0,
\end{align*}
after substituting $\delta a$ with its one-loop value as in \eqref{eq:counterterm}. Clearly, $v$ still minimizes the quantum effective action, it can be checked that the second derivative is positive.

\section{Vanishing of the tadpoles for $\braket{O}$ in the gauge-invariant formulation}\label{apptad}
The leading order contributions to the one-point function of $O$ are given in FIG.~\ref{tadpolelintransfo}.
\begin{figure}
	\centering
	\begin{tikzpicture}
		\begin{feynman}
			\vertex (x1) {\(O\)};
			\vertex [right= of x1](y1);
			
			\vertex [right=2cm of y1](x2) {\(O\)};
			\vertex [right= of x2](y2);
			
			\vertex [right=2cm of y2] (x3){\(O\)};
			\vertex [crossed dot,right= of x3] (y3){};
			\diagram*{
				(x1)--[edge label'=\(0\)](y1), 
				y1--[boson,edge label'=\(q\),out=-45, in=45, loop, min distance=2cm]y1,

				(x2)--[edge label'=\(0\)](y2), 
				y2--[edge label'=\(q\),out=-45, in=45, loop, min distance=2cm]y2,
				
				(x3)--[edge label'=\(0\)] (y3),
				()
			};
		\end{feynman}
	\end{tikzpicture}
	\caption{Feynman diagrams with one external $O$-field.}\label{tadpolelintransfo}
\end{figure}
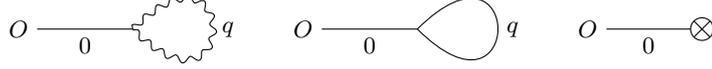
\begin{align*}
	\expval{O(x)}^{\text{1-loop}}_1&=-\frac{1}{2}\frac{1}{m_h^2}2e^2v\int \frac{d^dq}{(2\pi)^d} \frac{\delta_{\mu\nu} (\delta_{\mu\nu} +\frac{q_\mu q_\nu}{m_A^2})}{q^2+m_A^2}=-\frac{1}{m_h^2}e^2 v (m_A^2)^{d/2-1}\frac{\Gamma(1-d/2)}{(4\pi)^{d/2}}(d-1),
\end{align*}
\begin{align*}
	\expval{O(x)}^{\text{1-loop}}_2&=\frac{1}{2}\frac{1}{m_h^2}(-3\lambda v)\int \frac{d^dq}{(2\pi)^d} \frac{1}{q^2+m_h^2}=-\frac{1}{m_h^2}\frac{3}{2}\lambda v (m_h^2)^{d/2-1}\frac{\Gamma(1-d/2)}{(4\pi)^{d/2}},
\end{align*}
\begin{align*}
	\expval{O(x)}^{\text{1-loop}}_{CT}&=-\frac{v^2}{m_h^2} (\delta \sigma) v.
\end{align*}
The full one-loop contribution to the expectation value of $O$ is therefore given by:
\begin{align*}
	\expval{O(x)}^{\text{1-loop}}=-\frac{v}{m_h^2}\Bigg(\frac{\Gamma(1-d/2)}{(4\pi)^{d/2}} \Big(e^2 (m_A^2)^{d/2-1}(d-1) + \frac{3}{2}\lambda (m_h^2)^{d/2-1}\Big) +v^2(\delta \sigma)\Bigg).
\end{align*}
When choosing $\delta \sigma$ as in \eqref{eq:counterterm}, we once again obtain the vanishing of the tadpole diagrams.

\section{Explicit expressions for the 1-loop \(O\) propagator diagrams}
\label{apx:explicit-diagrams}
\begin{align*}
	\expval{O(p)O(-p)}^{\text{0-loop}}_{\ref{diag:trivial}}=\frac{v^2}{p^2+m_h^2},
\end{align*}
In the following expressions we use the notation \(H(m_1^2,m_2^2,p^2)=x(1-x)p^2+(1-x)m_1^2 +x m_2^2\).
\begin{align*}
	\expval{O(p)O(-p)}^{\text{1-loop}}_{\ref{diag:sunset-scalar-photon-photon}}&=\frac{1}{2}\Big(\frac{v^2}{p^2+m_h^2}\Big)^2\Big(\frac{8}{v^4}\Big)^2\Big(\frac{e^2v^4}{4}\Big)^2 \int \frac{d^dq}{(2\pi)^d} \Big(\frac{\delta_{\mu\rho}+\frac{q_\mu q_\rho}{m_A^2}}{q^2+m_A^2}\Big)\Big(\frac{\delta_{\nu\sigma}+\frac{(q-p)_\nu (q-p)_\sigma}{m_A^2}}{(q-p)^2+m_A^2}\Big)\\
	&\Big(\frac{q^2}{m_A^2}P_{\mu\nu}(q)+\frac{(q-p)^2}{m_A^2}P_{\mu\nu}(q-p)+\delta_{\mu\nu}\Big)\Big(\frac{q^2}{m_A^2}P_{\rho\sigma}(q)+\frac{(q-p)^2}{m_A^2}P_{\rho\sigma}(q-p)+\delta_{\rho\sigma}\Big)\\
	&=\Big(\frac{1}{p^2+m_h^2}\Big)^2 \int_0^1dx \frac{\Gamma(2-d/2)}{(4\pi)^{d/2}}\Big(2(d-1)m_A^4+2p^2m_A^2+\frac{p^4}{2}\Big)H(m_A^2,m_A^2,p^2)^{d/2-2}\\
	&+\frac{\Gamma(1-d/2)}{(4\pi)^{d/2}}\Bigg(-8m_A^4(d-1) +m_A^2p^2(4(d-2)+\frac{4}{d}-1)\Bigg)(m_A^2)^{d/2-2},
\end{align*}
\begin{align*}
	\expval{O(p)O(-p)}^{\text{1-loop}}_{\ref{diag:sunset-scalar-scalar-scalar}}&=\frac{1}{2}\Big(\frac{1}{p^2+m_h^2}\Big)^2\int \frac{d^dq}{(2\pi)^d} \frac{p^2+q^2+(q-p)^2}{q^2+m_h^2}\frac{p^2+q^2+(q-p)^2}{(q-p)^2+m_h^2}\\
	&=\Big(\frac{1}{p^2+m_h^2}\Big)^2\int_0^1dx\Bigg[\frac{\Gamma(1-d/2)}{(4\pi)^{d/2}}\Bigg(\Big(3p^2-4m_h^2\Big) (m_h^2)^{d/2-1}\Bigg)\\
	&+\frac{\Gamma(2-d/2)}{(4\pi)^{d/2}}\Big(\frac{p^4}{2}-2m_h^2p^2+2m_h^4\Big) H(m_h^2,m_h^2,p^2)^{d/2-2}\Bigg],
\end{align*}
\begin{align*}
	\expval{O(p)O(-p)}^{\text{1-loop}}_{\ref{diag:snail-scalar-scalar}}&=-2\Big(\frac{1}{p^2+m_h^2}\Big)^2\int \frac{d^dq}{(2\pi)^d} \frac{2(p^2+m_h^2)+2(q^2+m_h^2)-4m_h^2}{q^2+m_h^2}\\
	&=\Big(\frac{1}{p^2+m_h^2}\Big)^2\frac{\Gamma(1-d/2)}{(4\pi)^{d/2}}\Big( -4p^2+4m_h^2\Big)(m_h^2)^{d/2-1},
\end{align*}
\begin{align*}
	\expval{O(p)O(-p)}^{\text{1-loop}}_{\ref{diag:snail-scalar-photon}}&=\Big(\frac{v^2}{p^2+m_h^2}\Big)^2\frac{1}{2}\int \frac{d^dq}{(2\pi)^d}(-\frac{48}{v^6})\frac{e^2v^4}{4}\frac{\delta_{\mu\nu}+\frac{q_\mu q_\nu}{m_A^2}}{q^2+m_A^2}\Big(2\frac{q^2}{m_A^2}P_{\mu\nu}(q)+\frac{2}{3}\delta_{\mu\nu}\Big)\\
	&+\frac{16}{e^2v^8}\Big(-2 p^2 \delta_{\mu\nu}+2p_\mu p_\nu \Big)\frac{e^2v^4}{4}\frac{\delta_{\mu\nu}+\frac{q_\mu q_\nu}{m_A^2}}{q^2+m_A^2}\\
	&=\frac{1}{(p^2+m_h^2)^2} \frac{\Gamma(1-d/2)}{(4\pi)^{d/2}}\Bigg(8m_A^4(d-1)-m_A^2p^2 \Big(4(d-2)+\frac{4}{d}\Big)\Bigg)(m_A^2)^{d/2-2},
\end{align*}
\begin{align*}
	\expval{O(p)O(-p)}^{\text{1-loop}}_{\ref{diag:tadpole}}&=\Big(\frac{v^2}{p^2+m_h^2}\Big)^2\cdot\frac{2(p^2+m_h^2)-2m_h^2}{v^4}\cdot \expval{O}^\text{1-loop}_c\\
	&=\frac{1}{(p^2+m_h^2)^2}\frac{\Gamma(1-d/2)}{(4\pi)^{d/2}}p^2(m_h^2)^{d/2-1},
\end{align*}
\begin{align*}
	\expval{O(p)O(-p)}^{\text{1-loop}}_{\ref{diag:ct}}&=v^2\frac{1}{(p^2+m_h^2)^2}(-a_1 p^2-a_2 m_h^2)+2\frac{v^2}{p^2+m_h^2}(a_2-2\frac{\delta \sigma}{\lambda})\\
	&+(-\frac{1}{\lambda}a_2+\frac{4}{\lambda^2}(\delta \sigma)-\frac{2}{\lambda^2}(\delta a))\\
	&=\frac{v^2}{(p^2+m_h^2)^2}\Big(m_h^2(-2\frac{\delta a}{\lambda})+p^2(-a_1+4\frac{\delta \sigma}{\lambda}-4\frac{\delta a}{\lambda})\\
	+&\frac{p^4}{v^2}(-\frac{a_2}{\lambda}+4\frac{\delta \sigma}{\lambda^2}-2\frac{\delta a}{\lambda^2})	\Big).
\end{align*}

\bibliographystyle{apsrev4-2}
\bibliography{Bib}

\end{document}